# Analysis of uncertainty in the surgical department: durations, requests, and cancellations

Belinda Spratt[1✉], Erhan Kozan[1], Michael Sinnott[2]

[1]*Queensland University of Technology, Brisbane, AU*

[2]*Princess Alexandra Hospital, Woolloongabba, AU*

| | |
|---|---|
| Belinda Spratt | Email: b.spratt@qut.edu.au |
| | Phone: +617 3138 3035 |
| Erhan Kozan | Email: e.kozan@qut.edu.au |
| | Phone: +617 3138 2308 |
| Michael Sinnott | Email: Michael.sinnott@staffandpatientsafety.org |
| | Phone: +614 1220 7574 |

Word Count: 3064 (excl. title page, references, and tables).

**Acknowledgements**

This research was funded by the Australian Research Council (ARC) Linkage Grant LP 140100394. Computational resources and services used in this work were provided by the HPC and Research Support Group, Queensland University of Technology, Brisbane, Australia.

**Competing Interests**

The authors declare they have no competing interests.

# Analysis of uncertainty in the surgical department: durations, requests, and cancellations

**Abstract**

***Background:*** Analytical techniques are being implemented with increasing frequency to improve the management of surgical departments and to ensure that decisions are well-informed. Often these analytical techniques rely on the validity of underlying statistical assumptions, including those around choice of distribution when modelling uncertainty. ***Objective:*** The objective of the research is to determine a set of suitable statistical distributions and provide recommendations to assist hospital planning staff, based on three full years of historical data. ***Methods:*** Statistical analysis has been performed to determine the most appropriate distributions and models in a variety of surgical contexts. Data from 2013 to 2015 was collected from the surgical department at a large Australian public hospital. ***Results:*** A lognormal distribution approximation of the total duration of surgeries in an operating room is appropriate when considering probability of overtime. Surgical requests can be modelled as a Poisson process with rate dependent on urgency and day of the week. It is found that individual cancellations can be modelled as Bernoulli trials, with the probability of patient, staff, and resource based cancellations provided herein. ***Conclusions:*** The analysis presented here can be used to ensure that assumptions surrounding planning and scheduling in the surgical department are valid. Understanding the stochasticity in the surgical department may result in the implementation of more realistic decision models.

**Key Question Summary**

*What is known about the topic?*

Many surgical departments rely on crude estimates and general intuition to predict surgical duration, surgical requests (both elective and non-elective), and cancellations.

*What does this paper add?*

This paper describes how statistical analysis can be performed to validate common assumptions surrounding surgical uncertainty. The paper also provides a set of recommended distributions and associated parameters that can be used to model uncertainty in a large public hospital's surgical department.

*What are the implications for practitioners?*

The insights on surgical uncertainty provided here will prove valuable for administrative staff who wish to incorporate uncertainty in their surgical planning and scheduling decisions.

## 1. Introduction

The surgical department is a highly uncertain environment. Non-elective patients arrive without warning, patients cancel, and equipment fails. Additionally, inappropriate estimation of surgical durations leads to either underutilisation or overtime.

At present, many surgical departments rely on intuition and surgeon estimates to predict surgical durations. Even experienced surgeons tend to have low accuracy in their prediction of surgical duration[1]. In this paper, the most appropriate distribution is determined to approximate the total duration of surgeries assigned to an operating room.

Accurate predictions of surgical requests from patients are essential for producing robust schedules. Typically, prediction of surgical demand is performed on a long term basis (e.g. Lee,

Park [2], Martin, Rice [3]). Here, the appropriate choice of distribution for predicting surgical requests on a day-to-day basis is analysed.

Surgical cancellations are a frequent occurrence in hospitals, and may be classified as staff, patient, or resource based. In this paper, estimates for the probability of surgical cancellations are provided, enabling administrative staff to make informed decisions regarding operating theatre planning and scheduling.

In this paper, a case study hospital with a large surgical department is analysed. This statistical analysis is intended to provide practitioners and surgical scheduling staff with insights into the inherent uncertainty in the operating room. The objective of this research is to determine the most suitable distributions for modelling uncertainty in the operating theatre of an Australian public hospital. This is done in the hope that these insights will prove valuable for administrative staff who wish to incorporate uncertainty in their surgical planning and scheduling decisions.

## 2. Methods

The statistical analysis presented here is based on a case study of a large Australian public hospital using three full years of historical data. Operating theatre records were obtained from 2013 to 2015. These records were de-identified and de-duplicated prior to analysis. Statistical analysis has been performed using the statistical software $R^{©}$ version 3.1.1.

### 2.1. Statistical Analysis

When determining which model best fit the data, the parameters corresponding to the maximum log likelihood estimation are calculated for each of the distributions considered. The Akaike Information Criterion (AIC) is then calculated to measure the relative quality of each model. Any specialties with fewer than 25 observations are disregarded.

To test whether the assumption of lognormal surgical durations holds, the data is first transformed by taking the natural logarithm of the surgical durations. The transformed data is then tested to determine if it is likely to belong to a normal distribution. Data is visually analysed using density plots and standard normal Q-Q plots. Shapiro-Wilk, Anderson-Darling, and Lilliefors normality tests are used to determine the suitability of the distribution.

Bootstrapping is applied using the MATLAB® software to determine the 95th percentile of surgical durations for different numbers of patients in each of the non-elective specialties. For each specialty, n surgeries are randomly sample (with replacement) from the set of surgical durations 10,000 times and calculate the 95th percentile for the total length of n surgeries. Using this method, the maximum number of patients that can be treated in a half-day or full day block is determined, such that no overtime will be required in 95% of cases.

When considering elective surgery requests, surgeries listed as elective from the hospital's database are used and merged by patient reference number. Here, only waiting lists collected on Mondays (in this example) and only patients waiting for under a week are used to approximate elective surgery requests. Chi-square tests are performed to determine whether patients are added to the list uniformly throughout the week. Two-sample t-tests are used to determine whether the number of non-elective surgeries is uniform across weekdays. Two-sample t-tests are chosen due to the number of observations in the dataset. When determining if surgery requests follow a Poisson process, exact multinomial tests are preferred over chi-square goodness-of-fit tests, due to the small number of observations.

## 2.2. Data Availability

Whilst the original data is not accessible due to confidentiality agreements with the case study hospital, a set of randomly generated data based on the analysis presented here is provided online [4].

## 3. Results

### 3.1. Surgical Durations

In this subsection, the surgical duration of patients at the case study hospital are considered. Spratt and Kozan [5] use a lognormal distribution approximation for the duration of surgeries when creating schedules that reduce the probability of overtime. The following statistical analysis validates the use of a lognormal distribution when attempting to reduce the probability of overtime, in the case of both elective and non-elective patients.

In order to determine which distribution is the best fit for the surgical durations, distributions including Gamma, Lognormal, Normal, Cauchy, Logistic, Student's t, and Weibull are considered. Elective and non-elective patients are considered separately, and the data is further split by specialty.

Of the elective specialties with at least 25 observations, the distribution with the minimum AIC is lognormal. The only exception to this is the Cardiac Surgical Unit, which appears to be more suited to a skewed t distribution. As a lognormal distribution produced models with the minimum AIC, the data is transformed and tested using a Shapiro-Wilk normality test. The null hypothesis is as follows:

$H_0$: The logarithm of surgical durations of elective patients of specialty is normally distributed.

According to the goodness-of-fit tests used, the assumption of lognormal surgical durations is inappropriate for approximately half of the elective specialties. This may be due to the large (n

> 100) and heterogeneous populations. The specialties considered, number of observations, p-value from a Shapiro-Wilk goodness-of-fit test, mean, and variance of surgical durations (in hours), and lognormal distribution parameters ($\mu$ and $\sigma^2$) are presented in Table 1 (elective patients) and Table 2 (non-elective patients). Table 1 and Table 2 include the results from all specialties, despite low sample sizes. Results from specialties with small samples should be treated with caution. The goodness-of-fit tests performed indicate that roughly half of the non-elective specialties are likely to have lognormally distributed surgical durations.

Whilst a lognormal distribution may not be well suited to modelling individual surgical durations (when grouping by specialty), it may be useful for approximating the distribution of the total duration of surgeries assigned to an operating room. Spratt and Kozan [5] determine the number of patients assigned to an operating room time block such that the probability of overtime is at most 5%. This is calculated using a lognormal approximation. A bootstrapping methodology applied to historical data provides the same maximum number of patients as the lognormal approximation.

### 3.2. Surgery Requests

#### 3.2.1. Elective Surgery Requests

In this subsection, the arrivals on the elective waiting list are analysed on a weekly basis. Table 3 includes the average number of elective surgery requests per week, for each urgency category and specialty.

In Australian public hospitals, elective patients are placed on waiting lists based on the recommended maximum time until their surgery. There are three urgency categories. Category one patients are the most urgent patients and should be seen within 30 days of being placed on the waiting list. Category two and three patients should be seen within 90 days and 365 days respectively.

It is unlikely ($p \ll 0.01$) that the elective surgery requests (either overall, or for each urgency category) are uniformly distributed throughout the week. It is more likely that elective surgery requests are uniformly distributed on Tuesday through Friday, with $p = 0.68$, $p = 0.13$, and $p = 0.24$ for categories one, two, and three respectively.

Mondays, Sundays and Saturdays appear to be the main source of variability in the data. This assumption is tested by grouping Saturday, Sunday, and Monday as one category in the Chi-Square Test to test for uniformly distributed elective surgery requests. By grouping Saturday, Sunday, and Monday together, elective surgery requests can be modelled through a uniform distribution for both category one ($p = 0.053$) and two ($p = 0.207$) patients. It is unlikely that a uniform distribution holds for surgical requests of category three patients ($p \ll 0.01$).

Thus, when modelling surgical requests, for category one and two patients it is assumed that the rate of elective surgery requests is constant throughout the week. For category three patients, it is assumed that the rate of elective surgery requests varies throughout the week, spiking on Monday morning, low for the weekend, and constant for Tuesday through Friday.

In addition to determining the average rate of arrivals, it is important to determine whether arrivals follow a Poisson process as arrivals are often assumed to be Poisson in queuing theory. For simplicity, only Tuesday to Friday on the first week of nine waiting lists are considered. This results in 36 samples. There is insufficient evidence against the hypothesis that category one and three elective surgery requests follow a Poisson process ($p > 0.05$).

Results from an exact multinomial test indicate that there is very significant evidence ($p \ll 0.01$) that category two elective requests do not follow a Poisson process on Tuesdays to Fridays.

However when the arrival rate is specified for each day, there is insufficient evidence ($p > 0.05$) to reject the hypothesis that surgical requests follow a Poisson process on Wednesdays through Fridays. As such, the arrival rate must be modified throughout the week.

### 3.2.2. Non-elective Surgery Requests

In this subsection, the frequency of non-elective surgeries is discussed, in order to better estimate the appropriate capacity to reserve for non-elective patients of each specialty. The expected number of non-elective surgeries per day varies largely based on surgical specialty.

There are also slight variations around the day of the week with many specialties performing fewer non-elective surgeries on the weekend. Specialties including Breast and Endocrine, Gastroenterology and Upper Gastrointestinal & Soft Tissue appear to perform a slightly higher number of non-elective surgeries on the weekend. This may be due to the availability of staff and the scheduling strategies implemented rather than a reflection of demand.

In twelve of the 21 specialties considered, two-sample t-tests showed insufficient evidence against the hypothesis that the number of non-elective surgeries performed is uniform across each weekday. As such, historical averages can be used to estimate non-elective demand (Table 3).

### *3.3. Surgical Cancellations*

One of the largest sources of uncertainty in the surgical department is cancellation. These cancellations can be categorised as patient, staff, or resource related cancellations. A list of cancellation descriptions and average weekly frequency are provided in Table 4.

Patient cancellations should be expressed as a proportion of total patients. Although, realistically, some patients may be more likely to cancel than others, here it is assumed that each patient has the same probability of cancelling. It is also assumed that patient cancellations are independent. Through analysis of historical data, it is found that a patient on the elective surgery waiting list will cancel approximately 7.45% of the time. It is estimated that approximately 2.88% of elective patients cancel on the day of surgery.

There are on average 1.08 patients cancelled per week due to equipment failure. This does not equate to 1.08 ORs failing per week. In the nine months of available cancellation data, cancellations occurred on 29 days due to equipment failure. It is assumed that on these days, a single OR is unusable. The probability that an OR breaks down and is unavailable for a day is approximately 0.54%.

In terms of staff cancellations, these can be categorised as expected and unexpected. Here, when estimating staff cancellations, the only cancellations considered are those that occur due to short-notice leave requests (e.g. sick leave). On each day there is a 0.26% chance that a scheduled surgeon will be away or unfit to perform surgery. There is a 0.06% chance that an anaesthetist will be on short notice leave, causing surgical cancellations.

From the above analysis, it can be seen that individual cancellations can be modelled using Bernoulli trials, where the total number of cancellations is modelled through a Binomial distribution. For example, a Bernoulli trial with $p = 0.0288$ can be used to estimate the probability that an elective patient will cancel on the day of surgery. A Binomial distribution with parameters $n$ and $p$ can be used to model the total number of day-of-surgery patient cancellations, where $n$ is the number of elective patients scheduled for the day and $p$ is the probability of an individual cancelling. Similar methodology can be implemented when modelling machine breakdowns and staff short-notice leave.

4. **Discussion**

It is common to assume that surgical durations are lognormally distributed [6-8]. Here, it is shown that this assumption is questionable in approximately half of the specialties at a large Australian public hospital. In order to avoid this limitation, it is recommended that a lognormal approximation is used for determining the 95$^{th}$ percentile of surgical durations, as non-parametric tests indicate that this is a valid approximation. As such, this analysis is meaningful when implementing models to reduce the probability of surgical overtime.

In analysing the frequency of surgical requests, it is found that the frequency of requests differed throughout the week. In the case of elective surgery requests, it is likely that administrative staff members are rarely available to enter data on the weekend and that additional elective surgery requests are made on the Monday as a result of this. Statistical tests indicate that it is unlikely that the number of non-elective surgeries is uniformly distributed across the week. Here, it is assumed that any variation in the number of non-elective surgeries performed is due to the availability of staff and operating room time as opposed to a discrepancy in the number of non-elective presentations.

Whilst expected outages, maintenance, and staff leave can be incorporated into schedules, unexpected disruptions can have a large impact on the surgical department. As such, the focus here is on unexpected cancellations. Unexpected surgical cancellations can be classified as patient, staff, or resource based. In estimating patient cancellations, it is found that approximately 2.88% of elective patients cancel on the day of surgery. By examining historical data for days on which surgical cancellations were made due to equipment failure, it is found that on a given day an OR has a probability of 0.54% of a breakdown that results in cancellations. A surgeon takes short-notice leave on a day that they are scheduled for surgery with a probability of 0.26%. If additional hospital records were available, it may be more suitable to analyse short-notice surgeon leave for each individual surgeon, rather than as an aggregate group.

It should be noted that a non-elective patient is a patient who presented at the ED and is able to have their surgery within 24 hours of presentation. Elective patients are those who had their surgery scheduled through the waiting list system, and those who presented at ED but are unable to have their surgery within 24 hours of presentation. This is a data limitation that requires further consideration.

As waiting lists are only correct at the time of printing, this data may be missing patients that have previously been added and cancelled. By performing statistical analysis only on patients waiting under a week, the risk of missing patient information is reduced. The analysis provided here should be used as a guide for other surgical departments. Administrative staff should verify that the assumptions presented in this paper are valid in their department.

## 5. Conclusion

As the surgical department is a highly uncertain environment, it is necessary to better understand the inherent stochasticity in order to produce robust surgical schedules. In this paper, statistical analysis of a large surgical department was presented. Focus was placed on the short-term prediction of surgical requests, the duration of surgeries, and patient cancellations.

It was found that a lognormal distribution is useful when modelling the durations of both elective and non-elective surgeries. A Poisson process can be used to model surgical requests, however parameters are dependent on day of the week and urgency. The probability of patient cancellations was calculated both prior to the day of surgery and on the day of surgery as 7.45% and 2.88% respectively. The probability that an OR will breakdown and be unavailable for a day is estimated at 0.54%. The probability that a staff member will require leave with short-notice is 0.26%.

The analysis provided here can be used to guide the decisions of hospital administrative staff and practitioners. Through this work, overtime can be reduced and required capacity for non-elective patients can be determined more accurately. Overall, this leads to more efficient use of the surgical department.

**Appendix: Tables**

**Table 1**: Elective surgery durations by specialty.

| Specialty | Count | p-val | Mean | Variance | $\mu$ | $\sigma^2$ |
|---|---:|---:|---:|---:|---:|---:|
| Acute Surgical Unit | 23 | 0.52 | 2.229 | 1.225 | 0.691 | 0.220 |
| Breast & Endocrine | 539 | 0.37 | 1.776 | 0.741 | 0.469 | 0.211 |
| Colorectal | 353 | 0.00 | 2.866 | 5.749 | 0.788 | 0.531 |
| Cardiac Surgical Unit | 834 | 0.00 | 4.431 | 2.307 | 1.433 | 0.111 |
| Dental Surgery | 4 | 0.32 | 1.173 | 0.405 | 0.031 | 0.258 |
| Ear, Nose & Throat 1 | 34 | 0.01 | 4.313 | 12.430 | 1.206 | 0.512 |
| Ear, Nose & Throat 2 | 29 | 0.28 | 2.432 | 6.924 | 0.501 | 0.775 |
| Ear, Nose & Throat 3 | 3 | 0.40 | 1.982 | 0.764 | 0.595 | 0.178 |
| Faciomaxillary | 332 | 0.06 | 1.623 | 1.523 | 0.256 | 0.456 |
| Gastroenterology | 2 | NA | 0.887 | 0.020 | -0.132 | 0.025 |
| Gynaecology | 4 | 0.86 | 0.554 | 0.010 | -0.607 | 0.032 |
| Hepato-Pancreato-Biliary | 352 | 0.00 | 3.000 | 3.388 | 0.939 | 0.320 |
| Liver Transplant | 6 | 0.83 | 1.063 | 0.249 | -0.038 | 0.199 |
| Neurosurgery | 447 | 0.00 | 3.144 | 4.166 | 0.970 | 0.352 |
| Ophthalmology | 1957 | 0.00 | 0.792 | 0.360 | -0.460 | 0.454 |
| Orthopaedic | 1721 | 0.00 | 2.156 | 2.095 | 0.582 | 0.372 |
| Plastic Surgery | 1600 | 0.00 | 1.416 | 2.129 | -0.014 | 0.724 |
| Renal Transplant | 562 | 0.00 | 1.537 | 0.959 | 0.259 | 0.341 |
| Upper GI and Soft Tissue | 513 | 0.00 | 2.194 | 3.457 | 0.515 | 0.541 |
| Urology | 1464 | 0.00 | 1.437 | 1.913 | 0.035 | 0.656 |
| Vascular Surgery | 526 | 0.01 | 2.535 | 1.982 | 0.796 | 0.269 |

Table 2: Non-elective surgery durations by specialty.

| Specialty | Count | p-val | Mean | Variance | $\mu$ | $\sigma^2$ |
|---|---|---|---|---|---|---|
| Acute Surgical Unit | 698 | 0.05 | 1.863 | 1.190 | 0.475 | 0.295 |
| Anaesthetics | 13 | 0.06 | 0.966 | 0.378 | -0.205 | 0.340 |
| Breast & Endocrine | 75 | 0.36 | 2.066 | 1.610 | 0.566 | 0.320 |
| Cardiac Surgical Unit | 220 | 0.01 | 3.286 | 4.293 | 1.022 | 0.335 |
| Cardiology | 3 | 0.14 | 0.867 | 0.037 | -0.167 | 0.048 |
| Colorectal | 152 | 0.00 | 1.962 | 1.994 | 0.465 | 0.417 |
| Ear, Nose & Throat | 132 | 0.78 | 1.425 | 0.915 | 0.168 | 0.372 |
| Faciomaxillary | 329 | 0.75 | 1.554 | 0.718 | 0.311 | 0.260 |
| Gastroenterology | 224 | 0.62 | 1.090 | 0.484 | -0.085 | 0.342 |
| Hepato-Pancreato-Biliary | 138 | 0.10 | 2.357 | 1.790 | 0.718 | 0.279 |
| Liver Transplant | 48 | 0.00 | 6.732 | 7.395 | 1.831 | 0.151 |
| Neurosurgery | 278 | 0.20 | 2.297 | 1.332 | 0.719 | 0.225 |
| Ophthalmology | 144 | 0.03 | 1.685 | 0.740 | 0.406 | 0.232 |
| Orthopaedic | 2054 | 0.00 | 1.991 | 2.057 | 0.480 | 0.418 |
| Plastic Surgery | 465 | 0.00 | 1.462 | 1.351 | 0.135 | 0.490 |
| Renal Transplant | 281 | 0.00 | 2.906 | 2.153 | 0.953 | 0.227 |
| Respiratory | 6 | 0.72 | 0.872 | 0.076 | -0.185 | 0.095 |
| Trauma | 38 | 0.95 | 4.148 | 3.960 | 1.319 | 0.207 |
| Upper Gastrointestinal | 87 | 0.03 | 1.910 | 1.545 | 0.471 | 0.353 |
| Urology | 200 | 0.00 | 1.068 | 0.464 | -0.105 | 0.341 |
| Vascular Surgery | 357 | 0.02 | 2.144 | 2.099 | 0.575 | 0.376 |

Table 3: Average weekly surgical demand by specialty.

| Specialty | Elective | | | | Non-Electives | Total Demand |
|---|---|---|---|---|---|---|
| | Cat 1 | Cat 2 | Cat 3 | Total | | |
| Cardiology | 3.81 | 13.49 | 1.39 | 18.70 | 3.41 | 22.10 |
| Ear, Nose & Throat | 3.64 | 3.77 | 0.40 | 7.81 | 2.66 | 10.47 |
| General | 23.59 | 19.78 | 5.14 | 48.51 | 36.72 | 85.23 |
| Gynaecology | 0.19 | 7.11 | 0.58 | 7.88 | 0.00 | 7.88 |
| Neurosurgery | 3.37 | 22.88 | 10.13 | 36.39 | 5.61 | 42.00 |
| Ophthalmology | 2.67 | 14.31 | 13.36 | 30.33 | 3.55 | 33.88 |
| Orthopaedic | 12.00 | 0.11 | 0.04 | 12.15 | 35.67 | 47.82 |
| Plastic Surgery | 25.74 | 22.90 | 3.63 | 52.27 | 8.21 | 60.49 |
| Urology | 27.57 | 35.84 | 43.68 | 107.10 | 4.59 | 111.69 |
| Vascular Surgery | 8.07 | 9.36 | 0.36 | 17.78 | 6.51 | 24.30 |
| **Total Demand** | 110.64 | 149.56 | 78.71 | 338.92 | 106.95 | 445.87 |

**Table 4**: List of surgical cancellations and average weekly frequency.

| Cancellation Description | Count |
|---|---|
| Anaesthetist unavailable – insufficient staff | 0.15 |
| Anaesthetist unavailable – on leave | 0.12 |
| Anaesthetist unavailable – urgent case | 0.07 |
| Consultant removed patient | 3.93 |
| Data entry error | 8.78 |
| Deceased | 0.27 |
| Doctor elected not to perform case | 7.47 |
| Doctor unavailable – insufficient staff | 1.95 |
| Emergency, treated as | 0.93 |
| Equipment failure/unavailable | 1.09 |
| Failed to attend – day of surgery | 1.60 |
| Failed to attend – preadmission appointment | 0.05 |
| Insufficient staff – nursing | 0.14 |
| Insufficient staff – other | 0.83 |
| List rearranged – case brought forward | 11.05 |
| List rearranged – priority case | 28.65 |
| No beds | 1.63 |
| No ICU beds | 0.69 |
| No longer requires treatment | 5.05 |
| No OT time | 7.98 |
| Patient cancelled booking | 14.94 |
| Patient did not wait | 0.53 |
| Patient could not be located | 0.53 |
| Patient requested to be removed | 1.78 |
| Surgeon unavailable – on leave | 6.76 |
| Surgeon unavailable – urgent case | 0.93 |
| Transferred to other facility | 0.27 |
| Treated elsewhere | 1.53 |
| Unfit for surgery – condition | 14.61 |
| Unfit for surgery – preparation | 6.99 |
| **Total** | 131.31 |